\documentclass[twocolumn,showpacs,amssymb,aps,nofootinbib,floatfix]{revtex4-1}
\usepackage{epsfig}
\newcommand{\be}{\begin{eqnarray}}
\newcommand{\ee}{\end{eqnarray}}
\newcommand{\ave}[1]{\left\langle #1 \right\rangle}
\newcommand{\order}[1]{ \mathcal{O} \left( #1 \right) }

\begin{document} \hbadness=10000
\topmargin -0.8cm\oddsidemargin = -0.7cm\evensidemargin = -0.7cm
\title{Elliptic flow fluctuations in heavy ion collisions and the perfect fluid hypothesis}
\author{Sascha Vogel, Giorgio Torrieri, Marcus Bleicher}
\affiliation{Institut f\"ur Theoretische Physik,
  J.W. Goethe Universit\"at, Max von Laue-Stra\ss{}e 1, 60438 Frankfurt am Main, Germany  
torrieri@fias.uni-frankfurt.de}
\date{March 12, 2007}  

\begin{abstract}
We analyze the recently measured $v_2$ fluctuation in the context of
establishing the degree of fluidity of the matter produced in heavy ion collisions.
We argue that flow observables within  systems with a non-negligible
mean free path should acquire a
``dynamical'' fluctuation, due to the random nature of each collision
between the system's degrees of freedom.
Because of this, $v_2$ fluctuations can be used to estimate the
Knudsen number of the system produced at the relativistic heavy ion collider (RHIC).
To illustrate this quantitatively, we apply the  UrQMD model, with scaled cross sections, 
to show that collisions at RHIC have a Knudsen number at least one
order of magnitude  below 
the expected value for an  interacting hadron gas.  Furthermore, we argue that the Knudsen 
number is also bound from below by the $v_2$ fluctuation data, because
too small a Knudsen number
would break the observed scaling of $v_2$ fluctuations due to the onset of turbulent flow.
We propose, therefore that $v_2$ fluctuation measurements, together with an understanding of the turbulent regime
for relativistic hydrodynamics, will provide an upper as well as a lower limit for the Knudsen number.  We also argue that an energy scan of $v_2$ fluctuations could shed light on the onset of the fluid regime.
\end{abstract}
\maketitle


One of the most widely cited news (both in the academic and popular
press) coming out of the heavy ion community concerns the discovery of a 
``perfect fluid'' in collisions of heavy ions at RHIC \cite{v2popular,whitebrahms,whitephobos,whitestar,whitephenix}. 
The evidence for this claim comes from the successful modeling of the anisotropic 
expansion of the matter in the early stage of the reaction by means of ideal 
 hydrodynamics \cite{heinz,shuryak,huovinen}. This
argument is compounded by the sensitivity of anisotropic
expansion to shear viscosity \cite{romatschke,teaney}.
The presence of a non-negligible shear viscosity, therefore, can  
be detected by a careful analysis of  anisotropic expansion data.

However, apart from this evidence for a small viscosity many fundamental properties
of the fluid are unknown. E.g. it is currently discussed whether the observed fluid is a 
strongly interaction Quark-Gluon Plasma (sQGP) \cite{sqgp}, a bound
state Quark-Gluon Plasma (bsQGP) \cite{bsqgp} or 
a (turbulent) Glasma \cite{glasma} with instabilities \cite{turb1,turb2}.  

The connection between theory and experiment rests mainly on a single observable for the 
anisotropic expansion, namely the elliptic flow coefficient $v_2$.
The parameter $v_2$ is the  second Fourier component of the azimuthal anisotropy  of 
the particle momenta given by \cite{v2orig,v2orig2}
\begin{equation}
v_2 \equiv \ave{ \rm{cos}[2(\phi-\Phi_{RP})]} \quad,
\end{equation}
where $\phi$ denotes the azimuthal angle of one outgoing particles and $\Phi_{RP}$ is the azimuthal angle of the
reaction plane. The angular brackets denote an average over all considered particles from all events.

It should be stressed that (differential) studies of the collective flow are among the 
earliest predicted observable to probe heated and
compressed nuclear matter \cite{Scheid:1974yi}. As the transverse flow is intimately connected
to the pressure gradients in the early stage of the reaction, it provides information on the equation 
of state (EoS) and might therefore be used to search for abnormal matter states and 
phase transitions \cite{Stoecker:1979mj,Hofmann:1976dy,Stoecker:1986ci}.

The  elliptic flow is of special importance, because it 
is ``self-quenching'' \cite{v2orig,Sorge:1996pc,v2orig2}: The angular pressure gradients creating the anisotropy 
extinguish themselves shortly after the start of the hydrodynamic evolution. Thus, the final
$v_2$ is insensitive to later stages of the  evolution, providing a key hole  to the hottest, best thermalized, and
possibly deconfined phase of the reaction.

In this paper we use the recently measured event-by-event fluctuations of $v_2$
\cite{loizides,sorensen} to further investigate the properties of the fluid created at RHIC energies.
The experimental data suggests that the $v_2$ fluctuations follow  the
fluctuations in initial eccentricity $\epsilon$
\begin{equation}
\label{idealv2}
\omega_{v_2}=\sqrt{\frac{\ave{(\delta v_2)^2}}{\ave{v_2}^2}} 
=\sqrt{ \frac{\ave{(\delta \epsilon)^2}}{\ave{\epsilon}^2}}\quad.
\end{equation}
This relation follows in a straight-forward fashion from the proportionality
between $v_2$ and the eccentricity inferred from ideal 
\cite{heinz,shuryak,huovinen} and viscous \cite{romatschke} boost-invariant
hydrodynamics
\begin{equation}
\label{betadef}
v_2 = \beta \epsilon
\end{equation}
where $\beta$ is approximately constant (Indeed, since $\epsilon$ is a small dimensionless parameter driving anisotropy, this relation can be understood simply in terms of Taylor expansion).    

As long as we are far away from the turbulent regime (on which we comment later), the deterministic nature of hydrodynamics,its applicability event by event and Eq. \ref{betadef} constrain \cite{knudsen} the effect hydrodynamic propagation has on initial state fluctuations to the form
\begin{equation}
\label{v2propto}
\left. \delta v_2 \right|_{initial} \sim \beta \delta \epsilon
\end{equation}
Explicit calculations have confirmed that this is the case for ideal hydrodynamics \cite{ebye}.
A  non-zero viscosity should not alter the proportionality, but just lower the value of $\beta$ \cite{knudsen}.

It is, however, surprising that initial conditions be the {\em only}
source of fluctuations.   If the system is treated as a collection of interacting particles, the
random nature of each interaction should add a \textit{dynamical}
component to the fluctuation of any flow variable, which depends not on the 
initial conditions (with which it is not correlated) but on the random nature of each microscopic collision
\begin{eqnarray}
\label{v2fluctterms}
\ave{(\delta v_2)^2} \simeq   \beta^2 \ave{(\delta
        \epsilon)^2} + \Delta_{dyn}^2
\end{eqnarray}
One can quantify the degree of perfection of the
fluid by the Knudsen number \cite{knudsen} defined as:
\begin{equation}
\label{knudef}
Kn=\frac{\lambda}{ L} 
\simeq \frac{N_{particles}}{N_{collisions}}\quad,
\end{equation}
with $\lambda$ the mean free path of the particles, $L$ the 
typical length scale of the system $N_{particles}$  the total number of particles and $N_{collisions}$ 
denoting the total number of interactions (soft and hard).  

If the Knudsen number is zero, the system becomes a perfect fluid.   In this case, flow observables are fully deterministic.  Hence, the probability of a $v_2$ at a certain time assuming a given eccentricity is a $\delta-function$.
\begin{equation}
P \left( v_2 | \epsilon \right) \sim \delta \left( v_2 - \beta \epsilon   \right)
\label{ideallimit}
\end{equation}
and hence $\Delta_{dyn}$ is zero.

In the limit of large numbers of collisions, correlations between collisions become weak, so the probability distribution in Eq. \ref{ideallimit} becomes
\begin{equation}
P \left( v_2 | \epsilon ,Kn \right) \sim \frac{1}{2 \pi \sigma} e^{- \left( v_2 - \beta \epsilon  \right)^2/2 \sigma^2}
\label{gaussian}
\end{equation}
where $\sigma(Kn)$ goes to 0 as the Knudsen number goes to zero.
Thus, it is sensible to Taylor expand around $Kn$, so
\begin{equation}
\label{nkdef}
 \Delta_{dyn}  \sim \alpha \sqrt{Kn} + \order{Kn^2}
\end{equation}
Note that the only ``small parameter'' here is the Knudsen number. 
All dependence on the nature of degrees of freedom and their
interactions (in particular, whether the particles interacting are hadrons or partons, what is the equation of state etc.) is encoded within the parameter $\alpha$, which by naturalness is of order unity.
In the case of a vanishing Knudsen number (the ideal hydrodynamic
limit) $\Delta_{dyn}$ should vanish. 

Such a scaling, apparent in Kinetic theory, can be also derived within hydrodynamics 
\cite{landaubook}:  Fluctuations in fluids include a thermal fluctuation term (irrelevant here since $v_2$ is defined in a way that makes it independent of random multiplicity fluctuations) and a dynamical autocorrelation of the energy-momentum tensor.
This autocorrelation  scales
linearly with the shear and bulk viscosity, which in turn depend
linearly on the mean free path \cite{landaubook} and the inverse of
the typical number of collisions per particle.   Subsequent developments
\cite{csernai,zubarev} have not altered these basic conclusions, which have also been compared to Boltzmann equation simulations \cite{mansour}.  
The latter comparison makes us confident of the ``universality'' of our scaling, since hydrodynamic fluctuations treated in \cite{landaubook,zubarev,mansour} concern systems where the Boltzmann equation fails (e.g. water).   

This is important, since the definition in Eq. \ref{knudef} looks more natural within a Boltzmann equation formulation, which is in turn based on the scattering approximation between interactions.
It is not clear whether this is a good approximation to use within RHIC.  While some groups have managed to bring models based on these assumptions in agreement with RHIC data \cite{xugreiner}, the appearance of fields and off-shell effects in the strongly coupled limit is not unreasonable \cite{cassing}.
This matter is complicated by the fact that everyday fluids\footnote{The usual linguistic usage is ``liquid'' for the definition based on compressibility, and ``fluid'' for the definition based on viscosity.  These two are often used as synonym in everyday English. } (where compressibility is typically large and correlated with viscosity due to Pauli blocking effects) are fundamentally different from ultra-relativistic ones (where the Pauli principle is not thought to be relevant, compressibility is bound by causality and typically unrelated to viscosity).  In the latter, the everyday definition of fluid vs gas (based on compressibility) becomes inapplicable, and a small viscosity alone is no guarantee that the Boltzmann equation approximation is not a good one.

Our definition of Knudsen number, however, is general enough to be independent of these considerations:  The Knudsen number is simply the ratio of a microscopic scale (where quantum randomness is important) to the macroscopic scale, in this case the total size of the system.    It is easy to see that, to leading order, the ratio of the two quantities must generally be $\sim \eta/(sTL) \sim Kn$ \cite{dirketa}.  Such a definition of the Knudsen number allows us to recover the scaling found in \cite{landaubook,csernai,zubarev} (derived with systems where the Boltzmann equation is {\em not} a good approximation, such as water, in mind).

It is therefore apparent that $\omega_{v_2}$ is a test for the
hypothesis that the system at RHIC is a ``perfect fluid'',i.e. a
locally thermalized system, where ``many'' particles undergo ``many''
collisions over a ``small'' fraction of the system's evolution.   
Potentially, this test is considerably more model-independent than a hydrodynamic analysis of $\ave{v_2}$,
since $\Delta_{dyn}$ scales directly with the Knudsen number, and all other
factors are of order unity.   

Deviations from this limit, including plasma instabilities, or
clustering, should therefore contribute fluctuations to
$v_2$ \cite{shuryakfluct} that can be probed by
comparison to the newly available experimental data. Interestingly, the ``opposite'' limit to hydrodynamics, a classical non-Abelian field such as the ``color glass condensate'' (CGC) \cite{cgc}, is also fully deterministic, and hence would exhibit $\Delta_{dyn}=0$. Just like ideal hydrodynamics, however, the absence of dynamical fluctuations in the CGC is an artifact of it being an effective description with a zero ``small parameter''.
 Here, the ``small parameter'' giving rise to fluctuations would be the inverse of the occupancy number of each quantum state.  
As the initial occupancy number at RHIC is 3-4 \cite{cgc}, diminishing to $\leq 1$ as the CGC melts, we expect large dynamical fluctuations also in a CGC prethermal stage, through calculating them is best left to a future work.


In this work, we  quantitatively assess the sensitivity of $\omega_{v_2}$ on $Kn$ by a string/hadron transport approach, the
UrQMD v2.3 model \cite{urqmd1,urqmd2}. To explore the different regimes, 
we rescale the total 
interaction cross sections by factors of $1/2$, $1$ and $3$ to vary the strength of 
the interaction.

Note that we are not using this rescaled UrQMD as a realistic model of the system, but rather as a ``toy model'' to study the scaling of $v_2$ fluctuations of the Knudsen number.
We believe this is an appropriate approach for the present study because the results can, 
in a straight-forward fashion, be converted into an estimate for the Knudsen number in 
heavy ion collision at RHIC. We expect that an analysis with partonic degrees of freedom will yield the same
{\em scaling} with the Knudsen number, and a quantitative result within the same
order of magnitude, the differences being encoded in the constant $\alpha$ of Eq \ref{nkdef}. 

As UrQMD is a quantum molecular dynamics simulation, the Knudsen number can be effectively ``measured'' by keeping track of the collisions during the system's lifetime, with $Kn \sim \ave{N_{particles}/N_{collisions}} \sim 0.6-1.5$.  This analysis parametrically agrees with an estimate following \cite{knudsen}, from 
 the ratio of the calculated 
elliptic flow to the hydrodynamic expectation.

 UrQMD also accounts for the expected non-flow
effects, as well as the fluctuations in the initial condition, that
also contribute to $\omega_{v_2}$ \cite{zhu}. For a general discussion of the $v_2$ analysis 
within this approach the reader is referred to \cite{Bleicher:2000sx,Petersen:2006vm}.

\begin{figure}[t]
\epsfig{width=8cm,clip=1,figure=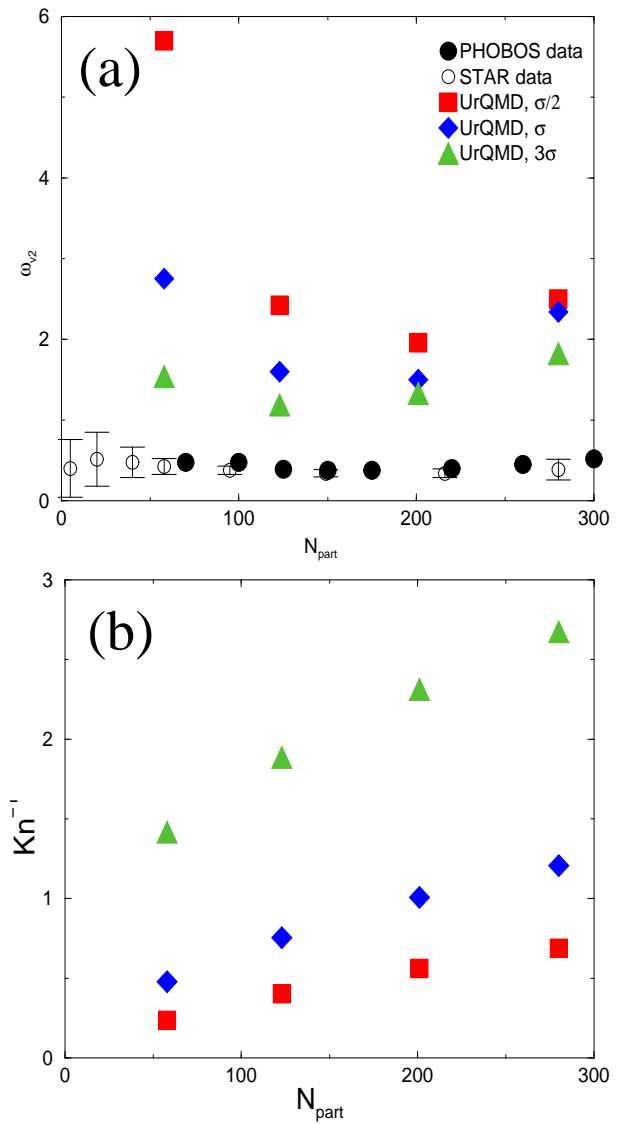}
\caption{\label{urqmd} (color online) UrQMD result for $\omega_{v_2}$ (panel (a) ) and the
Knudsen number (panel (b)) as a function of the number of participants for Au+Au reaction 
at $\sqrt s_{NN}=200$~GeV.  The data are taken from \cite{loizides,sorensen}.}
\end{figure}

The results of the present calculations are shown in Fig. \ref{urqmd}. 
As can be seen, $\omega_{v_2}$ and $Kn$ have the expected qualitative dependence on the
over-all scaling parameter:  As the factor used to rescale
interaction cross sections increases, $\ave{v_2}$ increases \cite{zhu} and
$\omega_{v_2}$ decreases. However, both $\ave{v_2}$ \cite{zhu}
and $\omega_{v_2}$ are well away from the data-points even if the
cross section is increased by a factor of three. Beyond the given increase,
we run into technical difficulties and grossly over-estimate the
total multiplicity of the system.  Hence, we are not able to
explore the scaling of the cross-section further than three times the
physical one within the present approach.

\begin{figure}[t]
\epsfig{width=8cm,clip=1,figure=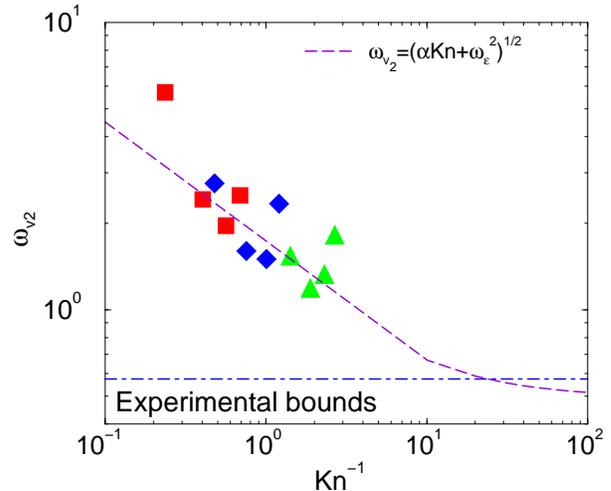}
\caption{\label{v2nk} (color online) Relationship between $\omega_{v_2}$ and the
  Knudsen number, plotted together with the Poissonian expectation.  The parameter
  $\alpha$ was fitted from the data.
See Fig. \ref{urqmd} for the legend}
\end{figure}

Fig. \ref{v2nk} shows the scaling of $\omega_{v_2}$ w.r.t. the Knudsen
number.  
The full line shows a fit of the calculations assuming a the additional fluctuations can be modeled by Eqs. \ref{v2fluctterms} and \ref{nkdef},
with $\alpha$ and $\beta$ are extracted from the fit.  While
$\alpha$ varies with the fundamental properties of the system even in the Poissonian limit, the fitted value is
sufficient for an order of magnitude
estimate of $Kn^{-1}$. 

  The scaling in Fig. \ref{v2nk} can be used to extrapolate to the inverse Knudsen number
needed to describe the elliptic flow fluctuations observed at RHIC.  Extrapolating the dependence of Eq. \ref{nkdef} to the upper experimental error bar for $\omega_{v2}$ yields a 
lower bound on the inverse Knudsen number on the order of a hundred. 
This estimate for the Knudsen number (or the potential scaling factor
for the cross section, as the Knudsen number is inversely proportional to the cross section) 
qualitatively agrees, with the opacity estimate derived from $\ave{v_2}$ 
using pQCD transport calculations \cite{molnar,Molnar:2004yh}, as well
as with the estimate obtained through transverse momentum fluctuations
\cite{gavin}. 

With the present calculation, we have established an  \textit{upper}
limit to the Knudsen number.  It is natural to ask what happens to
$\omega_{v_2}$ as the Knudsen number goes to zero, and the system  is
more closely approximated by ideal hydrodynamics.

The Knudsen number is related to another well known number in hydrodynamics, namely Reynold's number.
Reynold's number is defined as 
\begin{equation}
Re = \frac{H L  \ave{v}}{\eta} = \frac{sTL \ave{v}}{\eta}\sim 3 \frac{ \ave{v}}{Kn}\quad,
\end{equation}
with $H$ being the enthalpy, $\eta$ being the viscosity, $s$ being the entropy 
density, $T$ denoting the temperature and $\ave{v}$ the typical flow velocity. Thus, a small Knudsen 
number goes hand in hand with an increase of Reynold's number.

\begin{figure}[h]
\begin{center}
\epsfig{width=8cm,clip=1,figure=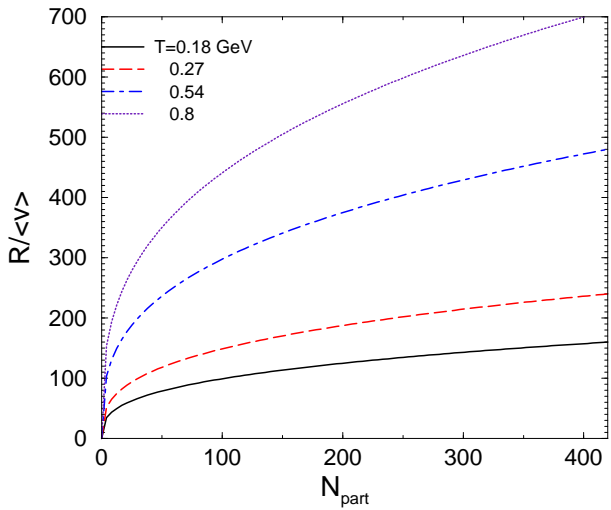}
\caption{\label{turbofig} (color online) The Reynolds number of a system the size of a
  collision between nuclei $A$, with the viscosity given by the
  conjectured ADS/CFT universal bound.}
\end{center}
\end{figure}

However, too high Reynolds numbers inevitably lead to instabilities of the hydrodynamic flow 
(the turbulent regime) and will add an additional source of fluctuations
to $\omega_{v_2}$, due to instabilities in the flow formation.
Estimating the Reynold's number for the present transport simulations leads to $Re \sim 1$.
However using the presently advocated ADS/CFT bound 
$\eta/s=1/4\pi$ \cite{unibound} leads to Reynolds numbers well into the $\sim 10^2$ in
the initial stages of the hydrodynamic evolution (Fig. \ref{turbofig}).  for $T = 200$~MeV, $L=10 fm$ and $\ave{v} \sim  1/\sqrt{3}$ (the speed of sound for a relativistic ideal gas), we have $Re \sim 100$,. 

Following \cite{landaubook} hydrodynamic instabilities will 
be present starting from $Re = 10-100$.  If  $Re>100-1000$ the flow will generally become turbulent \cite{landaubook}, although the onset of turbulence will also depend on the boundary conditions: the larger the bluffness of a layer of fluid (defined by $\ave{\left |u^i \times \vec{dA} \right|}$, where $u^i$ is the flow vector and $\vec{dA}$ the layer surface element) , the less Reynolds number is required for the onset of turbulence.     For a compressible fluid expanding from an ``almond-like'' shock, the last condition is likely to be satisfied close to the ``edges'' of the almond, provided that compressibility does not quench the onset of turbulence (the last question is not conclusively settled,
 through  recent evidence \cite{compress1,compress2} suggests that adding compressibility does not significantly change the critical Reynolds number for the onset of turbulence). 

Thus, we are led to conclude that, below a
certain critical Knudsen number, the scaling in Eq. \ref{v2fluctterms}
should break down and $\omega_{v_2}$ should increase significantly above
the ``ideal'' Eq. \ref{idealv2} value: While $\Delta_{dyn}$ would continue to decrease with decreasing
Knudsen number, $\omega_{v_2}$ would not anymore scale with
$\ave{(\delta \epsilon)^2}/\ave{\epsilon}$ but the initial fluctuation would be
amplified by the turbulent evolution.  If $\tau$ is the lifetime of the system and $\tau_0$ the timescale of the evolution of turbulence, $\omega_{v_2}$ in a turbulent fluid would scale as an exponent of an approximately power-law function of the initial volume (i.e. the number of participants)
\begin{equation}
\label{turbogrowth}
\omega_{v_2} \sim \left. \frac{\ave{(\delta
 v_2)^2}}{\ave{v_2 }}\right|_{\tau\ll\tau_0} e^{\tau/\tau_0} \sim  \frac{\ave{(\delta
 \epsilon)^2}}{\ave{\epsilon}} \exp \left[ N_{part}^\kappa \right]
\end{equation}
From causality (the time it takes for rarefaction waves to travel across the system, $\sim \mathrm{size}/c_s \sim N_{part}^{1/3}/c_s$,where $c_s$ is the speed of sound),  it can be deduced that $\kappa \simeq 1/3$.

It is apparent from Fig. \ref{urqmd} that such a scaling is
\textit{not} observed in the experimental data \cite{loizides,sorensen}, so the viscosity of
the system created at RHIC is high enough to place it out of the
turbulent regime.
This sustains the argument that the mere observation of a well-defined $\ave{v_2}$ places
a lower constraint on viscosity because it signals that the system
is not in a sufficiently turbulent regime. 

The implications of this statement on the closeness of the fluid
created at RHIC to the ADS/CFT viscosity bound are still not clear.
It is difficult to make a more precise estimate since
the turbulence in the system produced in heavy ion collisions has not
as yet been studied (for first attempts with QCD transport approaches, the reader 
is referred to \cite{turb1,turb2,Dumitru:2005gp,Strickland:2007fm}).  

In one dimension, the stability of boost-invariant dynamics (the boundary condition used,either exactly or in approximate form for simulations at RHIC energy) has been thoroughly studied.
Boost-invariant evolution was found to be generally stable at the early stages, where $v_2$ forms \cite{kouno,florkstab} (through instabilities could play a big role during freeze-out \cite{stab1,stab2}).  This leads us to think that if the system does have an early turbulent stage, $\tau_0$ remains long compared to its duration.
This conclusion is however bound to change within full 3D hydrodynamics,
especially if the system is not to a good
approximation boost invariant, as recent initial state calculations suggest \cite{Dumitru:2005gp,Strickland:2007fm}.

Thus, before a quantitative answer to these questions can be given,
a transport or hydrodynamic model capable
of modeling turbulence at the scale of heavy ion
collisions, and hence of inferring a quantitative value of $\tau_0$ in
Eq. \ref{turbogrowth}, is necessary.
Up to now, the only known calculation of the Reynolds number and the onset of turbulent flow 
in heavy ion collision has been done in Ref. \cite{Baym:1985tn}.

The onset of turbulence could be signaled experimentally
by a widening of $\omega_{v2}$ and a change of its dependence on $N_{part}$ from constant to exponential scaling as per Eq. \ref{turbogrowth}.
Thus, combined with the data in the non-turbulent regime, analyzed using the
ansatz of Eq. \ref{v2fluctterms}, the experimental measurement of
$\omega_{v_2}$ in a wide range of energies and system sizes can yield a lower as well as an upper limit of $Kn$.

\begin{figure}[t]
\begin{center}
\epsfig{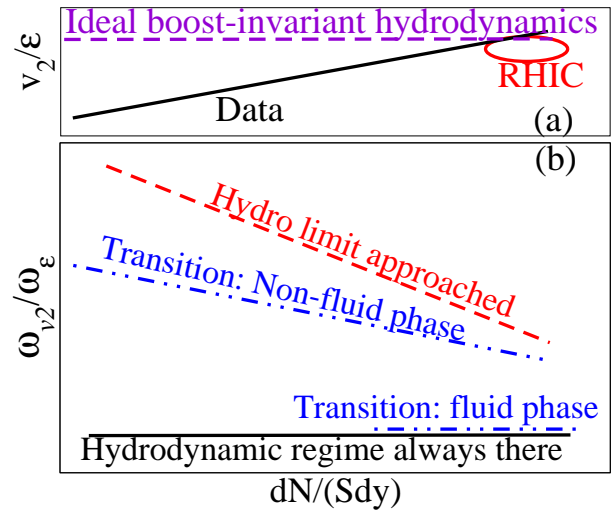}
\caption{\label{v2fluctscaling} (color online) A qualitative plot (panel (b)) showing the scaling of the $v_2$ fluctuations in the three scenarios suggested by the scaling of $v_2$ across energy and system size \cite{uniscaling1,uniscaling2,uniscaling3,uniscaling4}, shown in panel (a).  The $x$-axis, the rapidity density normalized by the overlap area, corresponds to the entropy density in the Bjorken hydrodynamic scenario.
If the hydrodynamic limit is smoothly approached with increasing system volume/lifetime the difference between the observed $v_2$ fluctuation and the initial estimate should go as the red dashed line. 
If the hydrodynamic limit indicates a transition between a viscous hadronic
gas and the sQGP, the scaling with $1/S dN/dy$ should be broken, with higher energy (``sQGP regime'') lower centrality events having a lower $v_2$ fluctuation than equivalent more central (``hadronic regime'') events.  This is indicated in the plot by the blue dot-dashed line.   Finally, a constancy of $\omega_{v2}/\omega_\epsilon$ might indicate that the ``hydro regime'' was actually with us all along, and only the initial conditions are responsible for the apparent rise in $v_2$.}
\end{center}
\end{figure}

Moreover, the energy dependence of $\omega_{v_2}$ could acquire a crucial phenomenological role in the light of the universal scaling seen
in $v_2/\epsilon$ \cite{uniscaling1,uniscaling2,uniscaling3} (panel (a) in Fig. \ref{v2fluctscaling}).  
The scaling variable is the multiplicity rapidity density normalized by the initial overlap surface $dN/(Sdy)$, chosen because it corresponds, in the boost-invariant picture \cite{bjorken}, to the entropy density divided by thermalization time.
One could interpret this scaling
as the approach to the ideal hydrodynamics limit as the initial density
become large.   If this interpretation is correct, the Knudsen number smoothly decreases as an inverse power of $dN/(Sdy)$, but has little sensitivity to the change in degrees of freedom at the phase transition \cite{uniscaling4}.
 Alternatively, it could be that the large flow observed at RHIC is indicative of a downward ``jump'' in the Knudsen number when the critical initial density needed to free partonic degrees of freedom is achieved.

The observation of $\omega_{v_2}$, and the excitation
function of $\omega_{v_2}/\omega_{\epsilon} $ could differentiate
between these scenarios (Fig. \ref{v2fluctscaling} panel (b)).  
If the system smoothly becomes more fluid at greater density, $\omega_{v_2}/\omega_{\epsilon} $ can be expected to decrease inversely with $dN/(Sdy)$ (smoothly if this is a continuous approach to hydrodynamics or
abruptly if the a transition to a more fluid regime is linked to a phase transition).   If fluidity is present in all systems to the same amount,  $\omega_{v_2}/\omega_{\epsilon} $ will stay constant across energies and system sizes. 

In conclusion, we have argued that the experimental observation of $\omega_{v_2}$ can provide unique 
information to estimate the Knudsen number  $Kn$, and hence to
to pin down  the perfection of the fluid
created in heavy ion collisions quantitatively. We have used a transport model to estimate a lower limit
of $Kn^{-1}$, and found that it is nearly two orders of magnitude below the value
needed to describe the $v_2$ fluctuations at RHIC.
We have also argued that the currently observed scaling of
$\omega_{v_2}$ should break in the turbulent regime, and hence the
measurement of $\omega_{v_2}$ potentially places an upper as well as a
lower limit on $Kn^{-1}$.   We have furthermore suggested that, in the light of these considerations, an energy and system size scan of $v_2$ fluctuations can shed light on the approach to the hydrodynamic regime.
However, before these limits can be quantitatively
ascertained, much more theoretical modeling and experimental
investigation is required.
 
The computational resources have been provided by the Center for Scientific Computing, CSC at Frankfurt University.
G.T. was
(financially) supported by the Helmholtz International Center for FAIR
within the framework of the LOEWE program (Landesoffensive zur
Entwicklung Wissenschaftlich-Ökonomischer Exzellenz) launched by the
State of Hesse.  We wish to thank H. Stoecker, Art Poskanzer, 
C. Greiner, M. Gyulassy, J. Rafelski,R. Venugopalan and S. Jeon for fruitful discussions.

\end{document}